\documentclass[12pt,fleqn]{article}
\usepackage{t1enc}
\usepackage[latin1]{inputenc}
\usepackage{graphicx}
\usepackage[dvips]{rotating}
\usepackage{flafter}
\begin{document}

\title{\bfseries Monte Carlo Simulation of Deffuant opinion dynamics with quality differences}

\author{Patrick Assmann\\Institute for Theoretical Physics, Cologne University, 50923 K\"oln, Germany}

\maketitle

\begin{abstract}
In this work the consequences of different opinion qualities in the Deffuant model were examined. If these qualities are randomly distributed, no different behavior was observed. In contrast to that, systematically assigned qualities had strong effects to the final opinion distribution. There was a high probability that the strongest opinion was one with a high quality. Furthermore, under the same conditions, this major opinion was much stronger than in the models without systematic differences. Finally, a society with systematic quality differences needed more tolerance to form a complete consensus than one without or with unsystematic ones.

\end{abstract}

{\bfseries Keywords}: Sociophysics, Monte Carlo simulations, Deffuant model, qualities.

\vspace{0.1cm}

\section{Introduction}
The computer simulation of opinion dynamics is an important part of sociophysics\cite{weid,moss1,schwei} and there exist a lot of different models and methods\cite{deff,deff2,hk,sz,gal,stauff}. The opinions represented by a number were randomly distributed among the simulated people (agents) and then some sort of opinion dynamics simulating a discussion is applied on the system. To our knowledge all these models assume no differences in the opinion quality. Every opinion has the same value.
But such an assumption seems not very realistic. Some opinions may have a higher quality due to a better argumentation structure or an ethical system which rewards some opinions e.g. by more social respect.\\
This examination now deals with the consequences of such differences in the opinion quality. The basic model is the discretized consensus model of Deffuant et al.\cite{deff2} with the agents located in a scale-free Barab\'asi-Albert network\cite{baralb1}.\\

\section{The different models}

In this study four different models were examined:

\begin{itemize}
\item{Model A: the basic model without any differences in the opinion quality; }
\item{Model B: the basic model with unsystematic quality differences;}
\item{Model C: the basic model with quality differences on an absolute scale;}
\item{Model D: the basic model with quality differences on a relative scale.}
\end{itemize}

\subsection{Model A: The basic model}
The model, which is used as basic model, is based on the consensus model of Deffuant et al.  To make the algorithm faster, the opinions were represented by integers\cite{stauff2} instead of real numbers on a continuous scale like in the original model of Deffuant. Therefore every agent \emph{i} has a number $S_i$ between 1 and \emph{Q} as opinion, where \emph{Q} is the maximum number of opinions.\\
Furthermore to use a realistic model of society instead of the simple 'everybody knows everybody' structure, the agents are located on the nodes of a directed Barab\'asi-Albert network\cite{baralb1}.\\
For this is a growing network, the construction process is dynamic. When $m$ is the number of outgoing connections of a node, the construction of the network starts with a core of $m$ nodes which are all connected to each other. Then, step by step, all other $N$ agents were added to the network. So you have a total number of $N+m$ agents. Every time a new node is added, it randomly chooses $m$ of the already connected nodes as neighbours. The probability to get linked to a node is proportional to the number of neighbours the node already has. So an agent with many 'friends' can get new 'friends' more easily. Here, we set $m=3$.\\
For the opinion dynamics two additional parameters are introduced, the confidence bound $\epsilon$ and the convergence parameter $\mu$.
 The interaction between the agents is pairwise. First the opinions difference $|S_i-S_j|$ of two discussion partners $i$ and $j$ is determined. If the difference is greater than the confidence bound $\epsilon$, nothing happens. If it is less than $\epsilon$, the discussion starts. For the opinions are integers, also $\epsilon$ should be an integer. To be independent from $Q$ a relative confidence bound $\epsilon_r$ is introduced which lies between 0 and 1: $\epsilon=[Q*\epsilon_r]$.\\
During a discussion both agents shift their opinion by the convergence factor $\mu$ towards each other. Here $\mu$ was set to $\sqrt{0.1}$. \\
\hspace*{3cm}\parbox{3.5cm}{$S_{ir}=S_i+[\nu*D]$\\$S_{jr}=S_j-[\nu*D]$} with \,  $\left\{ \begin{array}{r@{\quad}l} \nu = +\mu \,{\rm for}\, S_i < S_j\\ \nu = -\mu  \,{\rm for}\, S_i > S_j \end{array} \right.$\\
So that at least a little progress is achieved in every discussion, the minimum opinion shift is set to 1. If the two opinions differ only by 1, one agent simply takes the opinion of the other agent with a probability of 0.5.\\
The opinions of the agents are updated in sweeps over the whole population in the order of their integration into the network. Every agent randomly chooses at its turn one of its $m$ neighbours as discussion partner. If there is no change in the opinions of the agents during 10 iterations, the opinion distribution is considered as stable and the opinion dynamics ends.
The Deffuant model with these modifications is henceforth denoted as basic model.\\

\subsection{The quality differences}

Now the models with different opinion qualities are presented. In all models the quality differences are created by different convergence factors $\mu$. So an agent shifts its opinion with regard to the convergence factor of its actual opinion. A small $\mu$ will cause only a little shift. The agent does not like to leave its opinion. Therefore you could say the opinion has a high quality. On the other side an opinion with a bigger $\mu$ causes a higher shift and has therefore a smaller quality.\\\\
{\bfseries Unsystematic quality differences: Model B}\\
In this model the qualities are randomly assigned to the opinions. Every opinion $S$ gets its own convergence factor $\mu_S$ with $0 < \mu_S \le \mu$.
This unsystematic assignment could take place e.g. if the quality differences arise from argumentation structures of the opinions. Some have better arguments than others.\\
\hspace*{3cm}\parbox{3.5cm}{$S_{ir}=S_i+[\nu_{S_i}*D]$\\$S_{jr}=S_j-[\nu_{S_j}*D]$} with $n=i,j$\,  $\left\{ \begin{array}{r@{\quad}l}\nu_{S_n} = +\mu_{S_n} \,{\rm for}\, S_i < S_j  \\ \nu_{S_n} = -\mu_{S_n} \,{\rm for}\,  S_i > S_j \end{array} \right.$\\\\

{\bfseries Systematic quality differences: Model C and D}

In model C und D the qualities are systematically assigned to the opinions. Opinion 1 is set to the highest quality. All other opinions are related to this opinion.
This could occure e.g. in a society with an ethical system or a codex of behavior.

Both models use a different scale.\\
{\bfseries Model C} has an absolute scale. That means every opinion has its own constant quality regardless of other opinions. The convergence factor $\mu_S$ rises linear with the number of the opinion $S$:  $\mu_S=(\mu/Q)*S$.
The shift algorithm is the same as in model B.\\
{\bfseries Model D} however has a relative scale. Opinion 1 again has the highest quality, but all other opinions get their quality in regard of the opinion of the actual discussion partner. No opinion has its own, constant convergence factor. The convergence factors of both opinions being involved in a discussion are specifically determined for every discussion. This works as following: in case of a discussion the two opinions must differ at least by 2. So one of the opinions has a higher quality as the other because it is nearer to opinion 1. This opinion gets a reduced convergence factor which depends on the distance of both opinions: $\mu(D)=(1-D/\epsilon)*\mu$. The further the 'bad' opinion is away from the 'good' one, the less influence does it have. For the 'bad' opinion the normal convergence factor $\mu$ is used. That leads to\\
\hspace*{3cm}\parbox{3.5cm}{$S_{ir}=S_i+[\nu_i*D]$\\ \\$S_{jr}=S_j-[\nu_j*D]$} with \,  $\left\{ \begin{array}{r@{\quad}l} $\parbox{3.5cm}{$\nu_i=(1-D/\epsilon)*\mu \\ \nu_j=\mu $}$  \,{\rm for}\,S_i < S_j \\\\ $\parbox{3.5cm}{$\nu_i = -\mu \\ \nu_j=-(1-D/\epsilon)*\mu$}$ \,{\rm for}\, S_i > S_j \end{array} \right.$.

\section{Results}

\begin{figure}[t!]
\centering
\includegraphics[width=9cm,angle=-90]{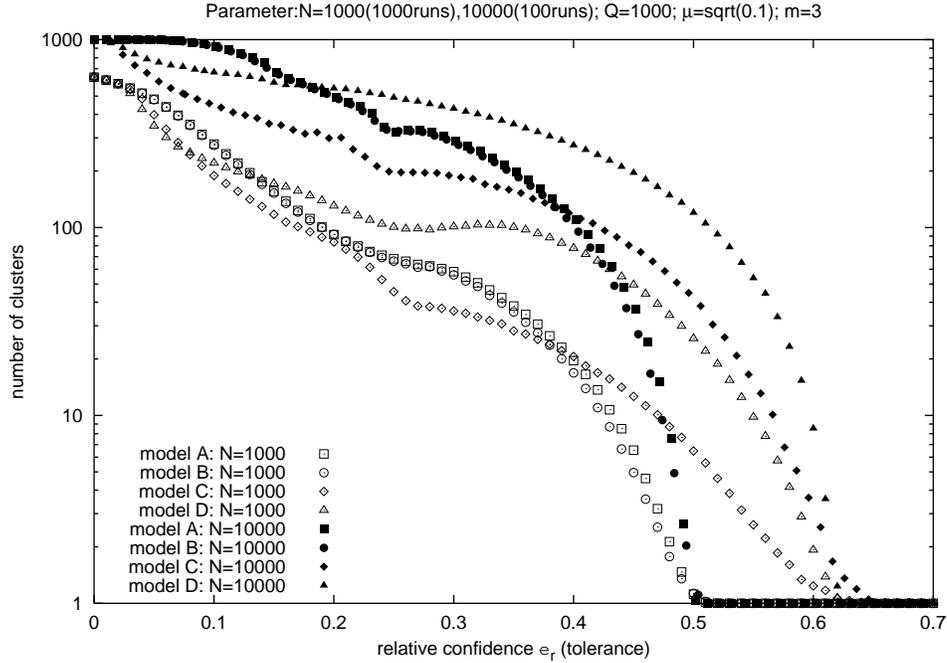}
\caption{\emph{Cluster number on a logarithmic scale for model A, B, C  and D each at $N=1000$ and $N=10000$.}}
\label {ass1}
\end{figure}

\begin{figure}[t]
\centering
\includegraphics[width=9cm,angle=-90]{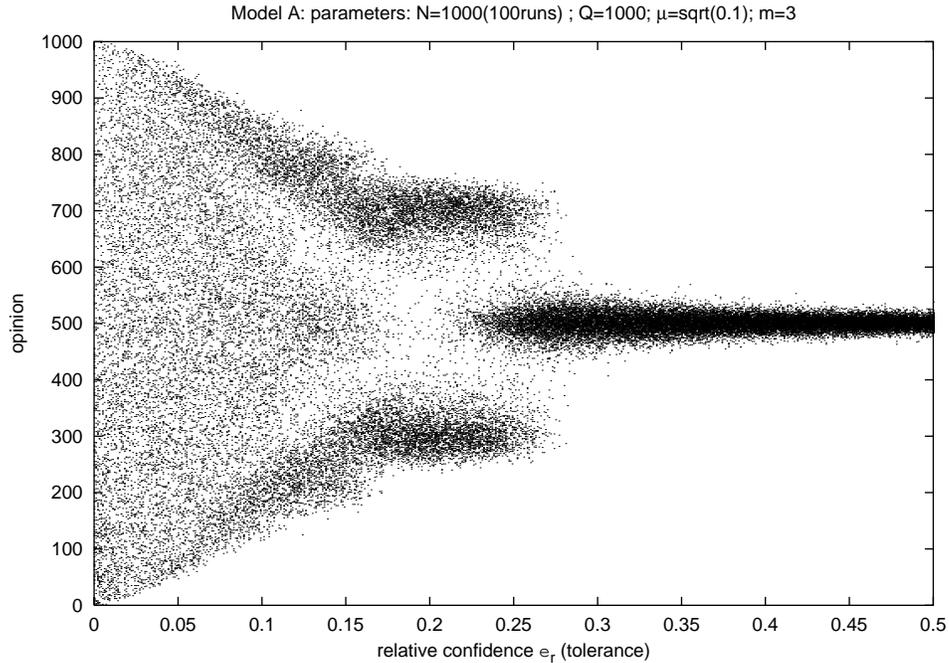}
\caption{\emph{Position of the maximum opinion for the basic model with N=1000}}
\label {ass2}
\end{figure}

\begin{figure}[t]
\centering
\includegraphics[width=9cm,angle=-90]{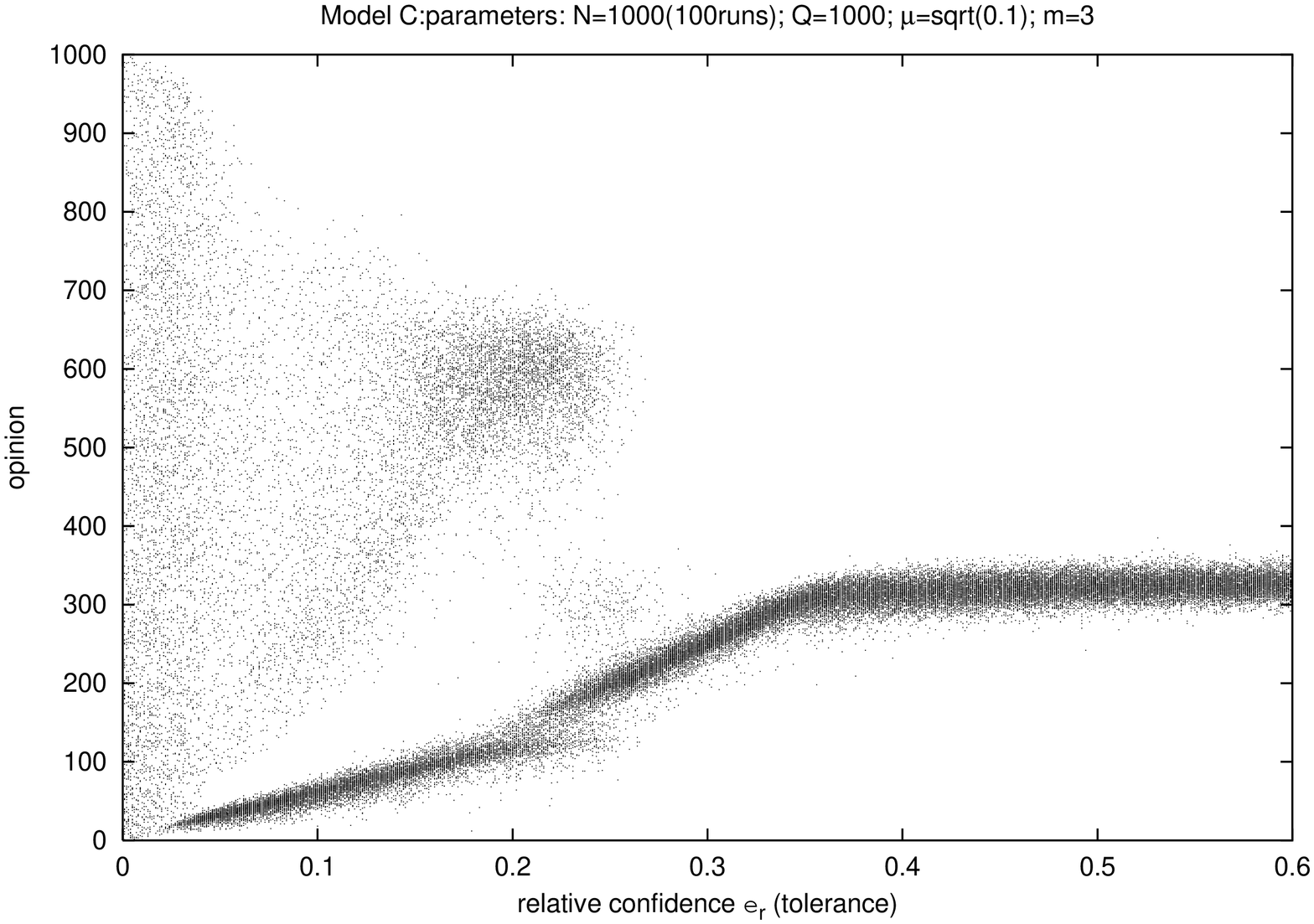}
\caption{\emph{Position of the maximum opinion for the model C with N=1000}}
\label {ass3}
\end{figure}

\begin{figure}[b!]
\centering
\includegraphics[width=9cm,angle=-90]{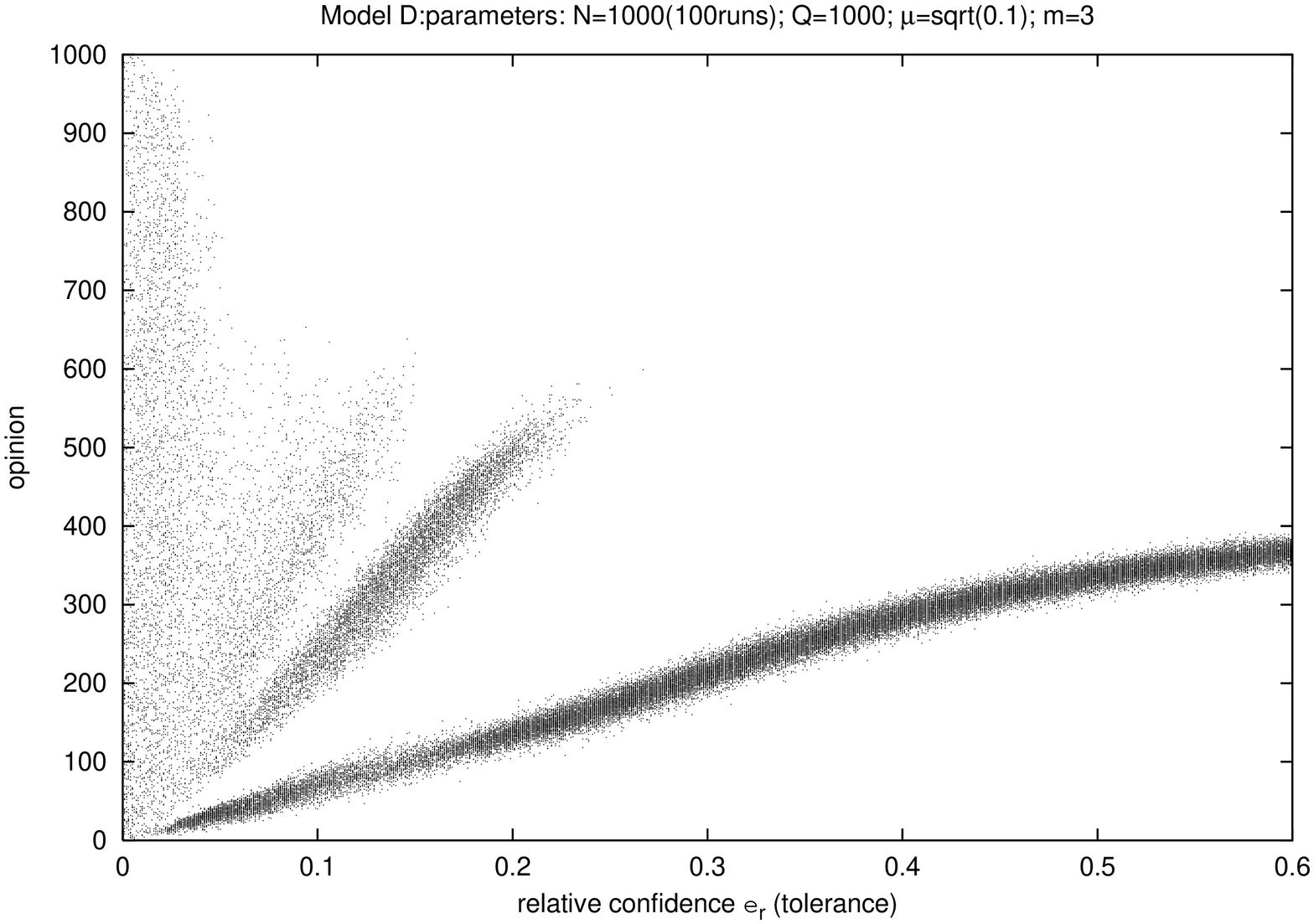}
\caption{\emph{Position of the maximum opinion for the model D with N=1000}}
\label {ass4}
\end{figure}

\begin{figure}[t]
\centering
\includegraphics[width=9cm,angle=-90]{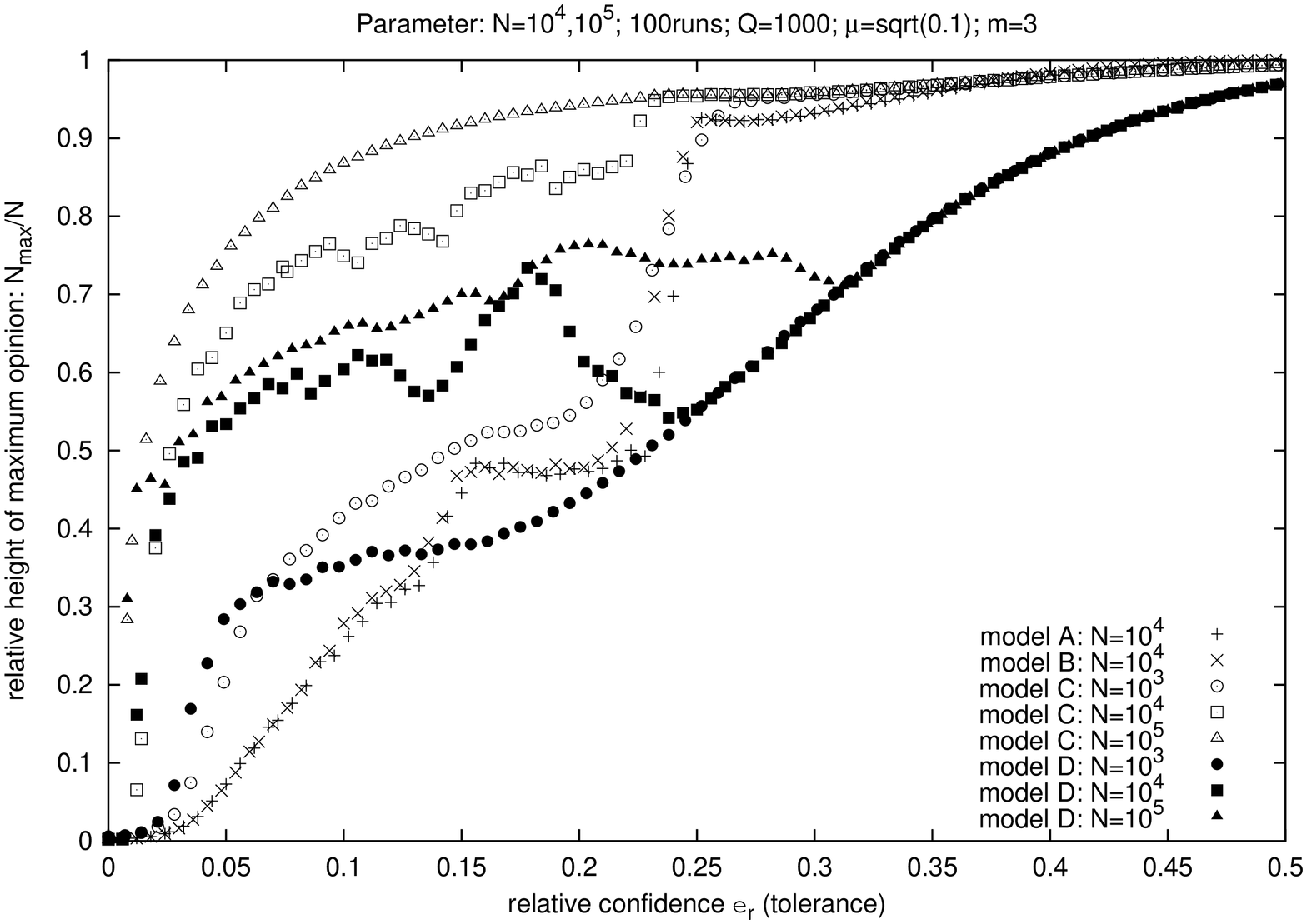}
\caption{\emph{Relative height of the maximum opinion for model A, B, C  and D each at $N=10^4$ and model C,D at $N=10^5$.}}
\label {ass5}
\end{figure}

One major point in analysing the models is the behavior of the maximum opinion. This is the opinion with the most clients at the end of the opinion dynamics. Both the position of the maximum opinion in the spectrum and its relative height were observed in dependence of $\epsilon$ and the size of the population. Furthermore the cluster number, that means the number of opinions which are occupied after the end of the opinion dynamics, is examined. All data points are averaged over at least 100 runs. Due to the way the quality differences were integrated in the standard model, the quality effects can actually only occur at a higher number of possible opinions. Therefore the number of opinions was set to $Q=1000$ in all simulations.\\

{\bfseries Model B} shows no major differences to the standard model (model A). The behavior of both models is qualitatively similar for all observed cases and there are only minimal quantitative differences. Fig.\ref{ass1} exemplifies this for the cluster number. There is nearly no difference between both curves. The same can be observed for the relative height of the maximum opinion in fig.\ref{ass5}. So the results of model B indicate that unsystematically distributed opinion qualities have no effect on the behavior of the discretized Deffuant algorithm.\\

{\bfseries Model C} and {\bfseries model D} show the same tendencies and are therefore treated together. The models without systematic qualities (A,B) reach the point of complete consensus at a confidence of $\epsilon_r\approx0.5$ (fig.\ref{ass1}). There only one opinion survives the discussion. All agents have the same opinion then. That this point of consensus is at $\epsilon_r\approx0.5$ was already examined in general for the continous Deffuant model \cite{santo}.
The models with systematic quality (C,D) now show a different behavior. Here the point of consensus is higher than 0.5. It is about 0.6 to 0.65.\\
 You can interpret the confidence bound as some sort of tolerance because it specifies the discussion range of an agent. One agent would not debate with another if their opinions differ too much. So, the confidence bound shows the tolerance of other opinions. In regards to that a society with systematic opinion qualities (ethical system) needs a higher tolerance of other opinions to reach a consensus than a society without such qualities.\\
To analyse the position of the maximum opinion, the final maximum position of every single run was plotted against the confidence. Every dot represents the strongest opinion in one run. Fig.\ref{ass2} shows this for the basic model with $N=1000$. As you can see the maximum opinions occur symmetrically to the center opinion. There is a high scattering of the positions at lower confidences. Here one cannot predetermine where the maximum will be. But for $\epsilon_r>0.3$ the maximum opinion appears only in a small bar around the center opinion. There the possible positions are contained sharply.
 This result is independent from the size of the population (examined sizes $N=100-10000$). The same distribution can be observed for model B.\\
In contrast to that, fig.\ref{ass3} shows the distribution of the maximum opinions for model C. Here again a high scattering occures at lower confidences, but there is a high probability for the maximum opinion to come out in the narrow bar at the 'better' side of the opinion spectrum.
For low confidence, this bar is at the very lower side of the spectrum and it shifts towards the center with increasing confidence. That means that the major opinion becomes less extreme with increasing tolerance of the society. The shape of this bar is nearly independent from the population size (only at larger numbers e.g. $N=10^5$ this bar becomes unsteady for lower confidences). However the shattering and the dark 'cloud' at $\epsilon\approx0.2$ vary with the number of agents. The higher the population size is, the smaller are these effects. Already at $N=10^4$ there is nearly no scattering and also the cloud has almost disappeared.\\
Fig.\ref{ass4} shows the position of the maxima for model D. Though the figure looks slightly different the systematics is the same as in model C. Instead of the "cloud' here a second 'branch' of possible opinion positions appears but with larger population sizes this also vanishes.\\
At last the relative height of the maximum opinion is examined. For all four models Fig.\ref{ass5} shows the fractional part of the population which holds this strongest opinion. Again there is almost no difference between the basic model (A) and the model with random opinion qualities (B). The shape of both curves is independent of the population size (and therefore, these curves were not plotted in this figure). Contrary to that, the models with systematic quality differences (C,D) do depend on the size of the population. Here, the number of agents holding the strongest opinion increases rapidly at lower confidences. Especially at larger $N$ the maximum opinion reaches the absolute majority very fast which means that more than half of all agents agree in one opinion already at low confidences.

\section{Summary}
In this study the effects of differences in the opinion qualities were investigated for the Deffuant consensus model. The simulations showed that there is almost no difference if the qualities are randomly distributed, i.e.\ independent of the opinion number, or if there are no opinion qualities at all.\\
If the qualities are not independent of the opinion numbers and are assigned systematically to these, the following results could be observed. There is a high probability that the opinion with the most clients is one with a very high quality. The probability rises with the population size. Furthermore this major opinion is already at low tolerance level very strong. This also increases with the number of agents. This is a nice effect because the qualities of the opinions are the same for all population sizes but they become more forceful for bigger societies. There seems to be some sort of herd behavior (stampede).\\
With growing tolerance, this major opinion grows even more, but at the same time it becomes less extreme. More different opinions can participate in the discussions and other points of view are presented. Therefore most agents find each other in a less extreme opinion.\\
To achieve a complete consensus there must be a higher tolerance level than in the models without a systematic quality distribution.\\
In general, all results are reasonable and can describe societies with an ethical system in a simple, but somehow adequate way. Therefore, the way of integrating different opinion qualities into the Deffuant model seems to be usable.

I am indebted to D. Stauffer for introducing me into the fascinating field of socio physics.

\end{document}